\documentstyle[twoside,fleqn,espcrc2]{article}

\newcommand\LG{Lan\-dau-Ginz\-burg\ }
\newcommand\ie{{\it i.e.} }
\newcommand\half{{1 \over 2}}
\newcommand\shalf{\coeff{1}{2}}

\def\advp#1{{\it Adv.\ Phys.}\ {\bf #1\/}}
\def\annp#1{{\it Ann.\ Phys.}\ {\bf #1\/}}
\def\cmp#1{{\it Commun.\ Math.\ Phys.} \ {\bf #1\/}}
\def\nup#1{{\it Nucl.\ Phys.} \ {\bf B#1\/}}
\def\plt#1{{\it Phys.\ Lett.}\ {\bf #1\/}}
\def\ijmp#1{{\it Int.\ J.\ Mod.\ Phys.}\ {\bf A#1\/}}
\def\jpa#1{{\it J.\ Phys.}\ {\bf A#1\/}}

\def\prl#1{{\it Phys.\ Rev.\ Lett. }\ {\bf #1}\/}

\def\rmp#1{{\it Rev.\ Mod.\ Phys.}\ {\bf #1}\/}

\newcommand\del{\partial}

\newcommand\Gminus{G_{-{1 \over 2}}^-}
\newcommand\Gplus{G_{-{1 \over 2}}^+}

\newcommand\Gmp{G_{-{1 \over 2}}^\mp}
\def\coeff#1#2{\relax{\textstyle {#1 \over #2}}
\displaystyle} 
\newcommand\inbar{\vrule height1.5ex width.4pt depth0pt}
\newcommand\IC{\relax\,\hbox{$\inbar\kern-.3em{\rm C}$}}
\newcommand\IR{\relax{\rm I\kern-.18em R}}
\font\sanse=cmss12
\newcommand\ZZ{\relax{\hbox{\sanse Z\kern-.42em Z}}}

\begin{document}
\pagestyle{empty}
\setcounter{page}{0}

\oddsidemargin 2cm
\topmargin  0cm

\rightline{\vbox{\baselineskip 12pt\hbox{hep-th/9512183}
\hbox{USC-95/027}}}
\vskip 3cm
\centerline{{\bf \Large \bf
Supersymmetric, Integrable Boundary Field Theories}}

\vskip 1.5cm
\centerline{{\bf \Large N.P. Warner}}
\vskip 1cm
\centerline{\it Physics Department, University of Southern
California}
\centerline{\it University Park,  Los Angeles, CA 90089--0484 USA}
\vskip 2cm
\noindent
Quantum integrable models that possess $N=2$ supersymmetry are
investigated on the half-space. Conformal perturbation theory is used
to identify some $N=2$ supersymmetric boundary integrable models, and
the effective boundary Landau-Ginzburg actions are constructed.
It is found that $N=2$ supersymmetry largely determines the boundary
action in terms of the bulk, and in particular, the boundary bosonic
potential is $|W|^2$, where $W$ is the bulk superpotential.
Supersymmetry is also discussed from the perspective of the affine
quantum group symmetry of exact scattering matrices, and exact $N=2$
supersymmetry preserving boundary reflection matrices are described.

\vfill
\noindent
Contribution to the proceedings of {\it ``Recent
Developments in Statistical Mechanics and Quantum Field
Theory,''} held at the ICTP, Trieste, April 10th  -- 12th, 1995.
\newcommand{\ttbs}{\char'134}
\newcommand{\AmS}{{\protect\the\textfont2
  A\kern-.1667em\lower.5ex\hbox{M}\kern-.125emS}}

\hyphenation{author another created financial paper re-commend-ed}

\title{Supersymmetric, Integrable Boundary Field Theories}

\author{N.P. Warner \address{Physics Department,
University of Southern California, \\
University Park,  Los Angeles, CA 90089--0484.}
\thanks{ Work supported in part by funds provided by the DOE
under grant No. DE-FG03-84ER40168.}}


\begin{abstract}
Quantum integrable models that possess $N=2$ supersymmetry are
investigated on the half-space. Conformal perturbation theory is used
to identify some $N=2$ supersymmetric boundary integrable models, and
the effective boundary Landau-Ginzburg actions are constructed.
It is found that $N=2$ supersymmetry largely determines the boundary
action in terms of the bulk, and in particular, the boundary bosonic
potential is $|W|^2$, where $W$ is the bulk superpotential.
Supersymmetry is also discussed from the perspective of the affine
quantum group symmetry of exact scattering matrices, and exact $N=2$
supersymmetry preserving boundary reflection matrices are described.
\end{abstract}

\maketitle

\oddsidemargin -4mm
\topmargin      -4mm
\pagestyle{plain}

\section{INTRODUCTION}

Apart from being an intrinsically interesting subject, the study of
conformal and quantum integrable models in two-dimensional systems
with a boundary also provides another avenue by which one can relate
integrable models to physically relevant systems and, in particular
to situations that are of interest in three- or four- dimensional
field theory. Some of these applications are discussed by other
speakers at this workshop.

In this paper, I will consider $(1+1)$-dimensional models defined
upon the half-line $(x<0)$ with a spatial boundary at $x=0$. The
field theory in the bulk ({\it i.e.} for $x<0$) will be required to
be $N=2$ supersymmetric, and either conformal or quantum integrable
(\ie possess higher spin conserved charges). The boundary conditions,
and boundary dynamics, will be chosen so as to preserve quantum
integrability along with as much supersymmetry as possible. In
general, for boundary models to be integrable there are stringent
constraints upon the bulk and boundary sectors (for example, see
\cite{GZam,Kondo,durham,itals}). As will be seen here, there
are $N=2$ supersymmetric boundary integrable models whose boundary
dynamics is quite non-trivial, with the boundary superpotential
determined by the bulk superpotential, but with a boundary mass scale
that is independent of the bulk mass scale.

There are several motivations for considering $N=2$ supersymmetric
boundary integrable models. First, if one has $N=2$ supersymmetry on
the plane or cylinder, then one can obtain quite a number of exact
quantum properties of such models via semi-classical analysis
\cite{OW,FMVW,WLNW,GenInd,EllGen,EGenNW}. One might hope that some of
these results can be generalized to the half-space. At a more formal
level, the supersymmetric models, and their topological counterparts,
provide the simplest examples of Coulomb gas methods, along with the
associated action of the affine quantum group on the soliton spectrum
and $S$-matrices. Once again, one would like to know how much of
these structures survive for quantum integrable models on the
half-space. Finally, from the point of view of higher dimensional
field theories, if one considers monopoles or strings in a
supersymmetric field theory, and treats them as impurity problems,
then the result will be supersymmetric $(1+1)$-dimensional boundary
field theories.

The starting point in this paper is $N=2$ superconformal models on
the plane. In section 2 I briefly review some pertinent facts about
conformal field theory and extended chiral algebras on the
half-space.  In section 3, I review the conditions under which
boundary and bulk perturbations of conformal models
lead to quantum integrable models,
and section 4 contains an analysis of the $N=2$ supersymmetric
boundary integrable models that can be obtained from perturbations of
the $N=2$ superconformal minimal series. In particular, it is shown
how families of such models can be obtained from the bulk
perturbations that lead to bulk superpotentials consisting of
Chebyshev polynomials. In section 5, effective boundary \LG actions
are constructed for these, and more general models.  It is
shown that if the bulk superpotential is $W(\phi)$, then the boundary
superpotential, $V$, satifies ${\partial V \over \partial \phi} = W$,
and hence the boundary bosonic potential is $|W|^2$.
Finally, section 6 contains a brief discussion of $N=2$ supersymmetry
preserving, exact boundary reflection matrices.

\section{CONFORMAL SYSTEMS WITH A BOUNDARY}

Consider a conformal field theory on the complete complex plane, and
suppose that the theory is symmetric between the left-moving and
right-moving sectors.  This means that every
left-moving operator, ${\cal O}(z)$, in the chiral algebra ${\cal
A}$, has a right-moving counterpart, $\widetilde {\cal O}(\bar z)$,
in the chiral algebra $\widetilde {\cal A} \equiv {\cal A}$.  As is
familiar in open string theory, the introduction of a boundary
requires that the left-movers and right-movers be locked together.
As a result, the two chiral algebras,   ${\cal A}$ and $\widetilde
{\cal A}$, become identified, producing a single copy,  $\hat
{\cal A}$, of  ${\cal A}$ in the system with the boundary.  Perhaps
the simplest way to think of this is as a generalized method of
images:  That is, ${\cal A}$-preserving boundary conditions require
\cite{Cardy}:
\begin{equation}
{\cal O}(z) \big |_{x=0}  ~=~ \widetilde {\cal O}(- \bar
z) \big |_{x=0} \ .
\label{lockO}
\end{equation}
One can then think of $\widetilde {\cal O}(- \bar z)$ as the analytic
continuation of ${\cal O}(z)$ into $x = {\cal R}e (z) > 0$.

In an $N=2$ superconformal theory the chiral algebra consists
of a spin-$1$ current, $J$, two spin-${3 \over 2}$ supercurrents,
$G^\pm$, and the energy momentum tensor, $T$.
For $N=2$ superconformal boundary
conditions, one requires:
\begin{eqnarray}
 J (z) \big |_{x=0}  & ~=~ & \ \widetilde J (- \bar z)
\big |_{x=0} \ , \nonumber \\
T (z) \big |_{x=0}  &~=~&  \ \widetilde
T (- \bar z) \big |_{x=0} \ , \nonumber \\
G^\pm (z) \big |_{x=0} &~=~&
\ \widetilde G^\pm (- \bar z) \big |_{x=0} \ .
\label{Abcond}
\end{eqnarray}
The result is a single $N=2$ superconformal algebra on the
half-space. The choice of pairing $G^\pm (z)$ with $\widetilde G^\pm
(- \bar z)$, or with $\widetilde G^\mp (- \bar z)$, is a matter of
convention, but changing this pairing will introduce a relative
negative sign in the pairing of $J(z)$ and $\widetilde J(z)$.

\section{BOUNDARY INTEGRABLE MODELS}

Given a boundary conformal model, there are two natural types
of massive perturbation.  The first is a (relevant) bulk
perturbation of the form:
\begin{equation}
\Delta A_{bulk} ~=~ g ~ \int_{{\cal R}e(z) < 0}
 ~\psi(z,\bar z)  ~d^2 z \ ,
\label{bulkpert}
\end{equation}
where $g$ is a coupling constant, and $\psi(z,\bar z)$ is generically
a sum of products of holomorphic and anti-holomorphic operators:
$\psi(z,\bar z) = \sum_j \psi_j(z) \tilde \psi_j(\bar z)$.  The
coupling $g$ introduces the bulk mass scale.

There are also boundary perturbations of the form:
\begin{equation}
A_{bdry} ~=~ \mu~ \int_{-\infty}^{\infty} ~\chi(t) ~dt
\ ,
\label{bdryact}
\end{equation}
where $\chi(t)$ is some operator defined upon the boundary at $x=0$.
To be relevant (or marginal), the operator $\chi$ must have dimension
less than (or equal to) one.  For a relevant boundary operator, the
coupling constant, $\mu$, introduces a boundary mass scale.  This
mass scale is {\it a priori} independent of the bulk mass scale.

\subsection{Conformal perturbation theory}

When there is no boundary, it is well known how to analyze whether a
bulk perturbation of a conformal field theory leads to an integrable
model \cite{Zamo,Toda}. The most direct method is to compute the
corrections to the conformal Ward identity $\bar \del {\cal O}(z) =
0$ for a current, ${\cal O}(z)$, in the presence of a perturbation,
$\psi(z,\bar z) = \psi(z) \tilde \psi(\bar z)$. The corrected
identity is of the form $\bar \del {\cal O}(z) = {\cal Y}$, and to
have a conserved current in the perturbed theory, one must have
${\cal Y} = \del {\cal Q}$, for some ${\cal Q}$. The form of the
correction ${\cal Y}$ is easily computed (to lowest order in $g$)
and  depends solely upon the operator ${\cal O}$, and the
represention of the chiral algebra to which $\psi$ belongs. (The null
vectors are usually crucial.)

If the system has a boundary, then there can be boundary
contributions to a conserved charge. The perturbative boundary
corrections to the bulk conservation laws can be computed much as in
the bulk theory \cite{GZam,NPW}. The result is once again a matter of
representation theory, and can be expressed in the following manner:
\begin{itemize}
\item [] If a bulk perturbation, $\psi$, generates a representation,
${\cal R}$, of the bulk chiral algebra, and leads to a quantum
integrable field theory in the system without boundary, then a
boundary operator $\chi$ in precisely the same representation, ${\cal
R}$, of the boundary chiral algebra will lead to an {\it integrable}
boundary field theory.  Moreover, the conserved charges of the
boundary theory are constructed from those of the bulk integrable
theory by pairing left-moving and right-moving currents as in
(\ref{lockO}), and adding boundary contributions to the
charge.
\end{itemize}
This result has been established only to first order in
perturbation theory. Under some circumstances one can argue that
there are no higher order corrections \cite{Zamo,GZam}. Even when
such arguments cannot be made, experience with integrable models in
systems without boundary suggests that first order
perturbation theory is usually sufficient to establish the existence
of conserved charges.

While the foregoing may seem straightforward prescription for
the construction of boundary integrable models, there can be
subtleties in obtaining the requisite boundary operator,
$\chi$. Specifically, in a given conformal model with conformal
boundary conditions, the boundary spectrum might not contain
a boundary operator in the required representation, ${\cal R}$
\cite{Cardy}.

\subsection{Spin-$1$ currents and topological charges}

There are two exceptions to the result described above.  The first
is trivial:  The holomorphic and anti-holomorphic parts, $J(z)$
and $\widetilde J(\bar z)$, of a spin-$1$ current in a conformal
model are separately conserved. If one makes a bulk perturbation,
then $J(z)$ and $\widetilde J(\bar z)$ give rise to a single
conserved $U(1)$ current if and only if the perturbing operator is
neutral with respect to some linear combination of the associated
left-moving and right-moving charges.  While this is obvious, the
computation in terms of conformal perturbation theory is
substantially different from the one employed in obtaining the
result above, and most
importantly, even though a bulk perturbation can preserve a $U(1)$
charge, the corresponding boundary perturbation may well destroy it
(even at first order).

The second exception is a little more subtle: if the bulk theory has
charges that are conserved at first order in perturbation theory, but
these charges do not commute with one another, then their commutators
will generate new charges that may not be conserved at the boundary,
even at first order in perturbation theory. This can happen with the
topological charges that appear in massive theories with extended
supersymmetry (for example, see \cite{OW,FMVW,WLNW,ABL}). These
topological charges appear in the anti-commutator of two
supersymmetries, and generically consist of integrals of total
derivatives of bosonic fields. As a result, the anti-commutators of
supersymmetry generators will pick up boundary terms that could break
the superalgebra, unless one does one of the following:
\begin{itemize}
\item [(i)]  Enforces bosonic boundary conditions that cause
the boundary terms to vanish.
\item [(ii)] Keeps the bulk conformally invariant, and only
perturbs on the boundary. There are no topological charges in a
bulk superconformal algebra.
\item [(iii)] Tries to compensate for the topological charge terms
by using boundary degrees of freedom, and making re-definitions
of the boundary supersymmetry and hamiltonian.
\end{itemize}
To illustrate possibility (iii), suppose that a bulk topological
charge term appears in the square of a supersymmetry generator and
gives rise to a boundary topological term ${\cal X}$. One can cancel
this term by introducing boundary fermions $b$ and $b^\dagger$, with
$\{b,b^\dagger\} = 1$, and then adding boundary correction of $b -
{\cal X} b^\dagger$ to the supercharge. Indeed, it will be seen later
that it is precisely this mechanism that enables one to construct
$N=2$ supersymmetry boundary integrable models with non-trivial
boundary interactions.

\section{SUPERSYMMETRY PRESERVING BOUNDARY PERURBATIONS}

I will, for simplicity, focus on the $N=2$ superconformal minimal
models  with $A$-type modular invariants, and whose central charge is
$c = 3k/(k+2)$.  The relevant bulk perturbations of these models
that lead to $N=2$ supersymmetric integrable field theories are
well known \cite{FMVW,FLMW,susyint}.  There are three distinct
such perturbations, but the one of importance here is the
Chebyshev perturbation.  That is, one perturbs the bulk using
\begin{eqnarray}
\lambda~\int~  \Gminus \widetilde \Gminus ~\phi_k^+
(z, \bar z) ~d^2z  ~+~  &&\nonumber \\
\ \   \bar \lambda~\int~\Gplus \widetilde \Gplus ~
\phi_k^- (z, \bar z) ~ d^2z \ , &&
\label{bulkp}
\end{eqnarray}
where $\lambda$ is a coupling constant, $\phi_k^+$ is the
chiral primary field with charge $q = \tilde q = {k \over k+2}$
and conformal weight $h = \tilde h = {k \over 2(k+2)}$, and
$\phi_k^-$ is the anti-chiral conjugate of $\phi_k^+$.
This perturbation leads to a massive $N=2$ supersymmetric
integrable model whose effective \LG potential is the Chebyshev
polynomial of degree $(k+2)$ (see, for example
\cite{NWTrieste}).  This may be thought of as the
$N=2$ superconformal analogue of the energy perturbation of the
ordinary minimal series.

Suppose that $\psi^\pm(t)$ are boundary operators that
transform in exactly the same representations as the
holomorphic operators $\Gmp \phi_k^\pm(z)$ under the action of
the $N=2$ superconformal algebra.  In particular this implies
that the operators $\psi^\pm(t)$ may be written
\begin{equation}
\psi^\pm(t) ~=~ \widehat \Gmp \Big(\hat\phi_k^\pm (t)\Big) \ ,
\label{psidefn}
\end{equation}
where $\widehat \Gmp$ are the supercharges of the boundary theory,
and $\hat \phi_k^\pm(t)$ are boundary operators in the same
$N=2$ superconformal representation as $\phi_k^\pm (z)$ and
$\phi_k^\pm (\bar z)$.

In order to use $\psi^\pm(t)$ in a boundary perturbation, it is
necessary to show that these operators are in the boundary spectrum
of some conformally invariant boundary condition. This is
straightforward, and is completely parallel with the analogous
situation in the Ising model. Let $\phi^+_1(z,\bar z)$ be the most
relevant chiral primary field (with $q = \tilde q = {1 \over k+2}$
and conformal weight $h = \tilde h = {1 \over 2(k+2)}$). This field
is the basic order parameter of the massive, bulk integrable model
\cite{ntwoLG,FMVW}. If we send this bulk field to the boundary, the
result is an operator in the fusion product of $\phi^+_1(z)$ with
itself. There are two fields in this fusion product: (i) $\phi^+_2$,
the chiral primary with $q = {2 \over k+2}$ and conformal weight $h =
{2 \over 2(k+2)}$, and (ii) the field $\widehat\Gplus \hat \phi^-_k$,
with $q = 1 - {k \over k+2} = {2 \over k+2}$ and conformal weight
$h = \half + {k \over 2(k+2)} = {k+1 \over k+2}$. If one starts with
the $N=2$ superconformal model with free boundary conditions in the
\LG formulation, then the expectation value of $\phi^+_2 =
(\phi^+_1)^2$ will vanish at the boundary, and one will get the
sub-leading operator $\psi^-(t) = \widehat\Gplus \hat \phi_k^- (t) $.
Similarly, one obtains $\psi^+(t) = \widehat\Gminus \hat \phi_k^+(t)$
by sending the anti-chiral conjugate, $\phi_1^- (z, \bar z)$, of
$\phi_1^+ (z, \bar z)$, to the boundary.

Conformal perturbation theory can now be invoked to show that (at
least to first order) one can obtain a boundary integrable model from
a simultaneous bulk perturbation using (\ref{bulkp}) and boundary
perturbation using (\ref{bdryact}) with:
\begin{equation}
\chi(t) ~=~ \nu ~\psi^+(t) ~+~ \bar \nu ~\psi^-(t) \ .
\label{chidefn}
\end{equation}
The constant, $\nu$, is a complex phase, with complex conjugate $\bar
\nu$. If there is no bulk perturbation then one can absorb this phase
$\nu$ into a re-definition of $\phi_k^\pm$. If there is a bulk
perturbation, then this freedom of re-definition can be used to
adjust the phase of $\lambda$ (in (\ref{bulkp})) or the phase, $\nu$.
Thus there are three independent couplings: the bulk mass scale,
determined by $|\lambda|$, the boundary mass scale, determined by
$\mu$ in (\ref{bdryact}), and the relative phase between
$\lambda/ \bar \lambda$ and $\nu/\bar \nu$.

It is well known that the bulk perturbation (\ref{bulkp}), in the
infinite domain, preserves all of the supersymmetry
\cite{FMVW,ABL,NWTrieste,PFKI,LNW}, and provides one with a massive
$N=2$ supersymmetric model. There are two ways of seeing the latter:
One can either explicitly verify the result using conformal
perturbation theory, or, more generally, one can appeal to the
theory of supersymmetric actions, which states that perturbations
that involve only top components of superfields (as does
(\ref{bulkp})) preserve the
supersymmetry. The same arguments can also be naively applied to the
boundary perturbation (\ref{chidefn}), with similar conclusions.
However, the massive bulk model has topological charges, and so the
supersymmetry variation of the bulk action generates boundary terms
via total derivatives. As described earlier, unless the bulk
is massless, the boundary conditions or dynamics need to be
fixed with care if one is going to preserve $N=2$ supersymmetry.
I will return to this issue
later, but first I think it useful to present a simple and fairly
complete example.

\subsection{An example: The sine-Gordon model}

The simplest example of an $N=2$ superconformal minimal model is
the $k=1, c=1$ minimal model that can be realized by a single free
boson compactified at the ``supersymmetric'' radius.  The
superconformal generators can be written in terms of the holomorphic
part of a canonically normalized boson, $\varphi(z)$, as:
\begin{eqnarray}
T(z) ~=~ -\coeff{1}{2}~ (\partial \varphi(z))^2 &; &
J(z) ~=~\coeff{i}{\sqrt{3}}~ \partial \varphi(z)\ ;
\nonumber \\
G^\pm(z)  ~=~  e^{\pm i \sqrt{3} \varphi(z)} \ . & &
\qquad \qquad
\label{susybose}
\end{eqnarray}
If $\tilde \varphi(\bar z)$ denotes the anti-holomorphic part of the
boson, then the $N=2$ superconformal boundary condition implies that
$\varphi(z)|_{x=0}  = \tilde \varphi(-\bar z)|_{x=0}$.  The order
parameter, and its conjugate,  are given by $\phi_1^\pm(z, \bar z) =
e^{\pm {i \over \sqrt{3}} (\varphi(z) + \varphi(\bar z))}$,
and as $x \to 0$ this
becomes $e^{\pm{2i \over \sqrt{3}} \varphi(z)} |_{x=0} = \Gmp
\phi_1^\pm(z)|_{x=0}$.  The bulk integrable perturbations are
$\Gmp \widetilde \Gmp \phi^\pm_1 =  e^{\mp {2i \over \sqrt{3}}(
\varphi(z) + \varphi(\bar z))}$.   We are thus dealing with the
boundary sine-Gordon theory described in \cite{GZam}, with action:
\begin{eqnarray}
\int_{-\infty}^{\infty} \int_{x<0}~ \big( \coeff{1}{2}
(\partial_t \Phi)^2 -  \coeff{1}{2} (\partial_x
\Phi)^2 \big) ~-~  \qquad && \nonumber \\  \qquad \qquad \qquad
{M \over \beta}~ \big[ cos \big(  \beta \Phi \big)
{}~-~ 1 \big] ~dx~dt ~~-~ && \nonumber \\  \qquad \qquad
{2m \over \beta}  ~  \int_{-\infty}^{\infty} ~ cos
\coeff{\beta}{2} \big(  \Phi -  \Phi_0 \big) ~dt  &&\ .
\label{bdrySG}
\end{eqnarray}
{}From this one finds that $\Phi$ satisfies the usual sine-Gordon
equation with boundary condition:
\begin{equation}
\partial \Phi \big |_{x=0}  ~=~ m~sin \coeff{\beta}{2}
\big(  \Phi -  \Phi_0 \big) \big |_{x=0} \ .
\label{sGbcs}
\end{equation}
The boson, $\Phi$, has the standard normalization of
sine-Gordon theory, for which the supersymmetry point
corresponds to $\beta^2 = {16 \pi \over 3}$.  The parameters
$M, m$ and $\Phi_0$ coincide with the three parameters
described above.  It is known that this model
is indeed integrable for arbitrary choices of $M, m$ and $\Phi_0$,
and so the boundary and bulk perturbations are simultaneously
and independently integrable.

\section{MANIFESTLY SUPERSYMMETRIC ACTIONS}

In the sine-Gordon model, the action is by no means manifestly
supersymmetric.  Indeed, the supersymmetry is obtained from
vertex operators and appears as an ``accident'' of the choice of
the coupling, $\beta$.  To obtain manifestly supersymmetric
actions for the $N=2$ superconformal minimal models, perhaps
the easiest way to proceed is to use the \LG approach in
which one looks for an effective field theory of the order
parameter $\phi_1$ \cite{ntwoLG}.  The idea here is to
generalize this to boundary field theories.

The first step is to make the elementary observation that
the boundary perturbations, defined by (\ref{psidefn}) and
(\ref{chidefn}), must be fermionic operators, and so there is no way
that they can be added by themselves to this action. The only option
is to introduce a dimensionless boundary fermions, $b$ and
$b^\dagger$, with $\{b,b\} = \{b^\dagger, b^\dagger\} = 0$, and
$\{b,b^\dagger\} = 1$.   One then considers
boundary perturbations of the form:
\begin{equation}
\nu~ b^\dagger~\psi^+ ~+~ \bar\nu~\psi^-~b \ .
\label{goodbpert}
\end{equation}
Note that if $\psi^\pm$ transforms under a $U(1)$ charge, then
one can arrange that the action preserve this $U(1)$ by
making $b$ and $b^\dagger$ transform appropriately. (It was
for this reason that I did not use a real fermion, $b$,
with $\{b,b\} = 1$.)

A more standard way to think of the boundary operators, $b$ and
$b^\dagger$, is as introducing a boundary spin degree of freedom
exactly as one does in the Kondo problem. Indeed, the quantization
rules for $b$ and $b^\dagger$ imply that they generate some (possibly
time dependent) representation of the gamma matrix algebra. In the
sine-Gordon model, the boundary interaction (\ref{goodbpert})
corresponds to coupling vertex operators to spin raising and lowering
operators.

For the sine-Gordon model, one can also see rather explicitly that
one should add these complex boundary degrees of freedom. One
considers the model as one approaches the limit of free boundary
conditions. As discussed in \cite{GZam}, any boundary degrees of
freedom will not disappear in the free limit, but they will simply
decouple from the bulk and will become massless boundary excitations.
These will appear as poles at $\theta = i \pi/2$ in the boundary
reflection matrix. One indeed finds such poles in all channels of the
reflection matrix. Since the soliton and the anti-soliton have
fermion numbers $\pm \half$, the off-diagonal poles in the reflection
matrix indicate that there will be boundary excitations that carry
fermion number $\pm 1$. The corresponding operators are $b$ and
$b^\dagger$.

\subsection{Non-trivial, supersymmetric boundary interactions}

The free $N=2$ superconformal model with a complex
boson and a complex Dirac fermion has central charge
$c=3$, and the (Euclidean) action is:
\begin{eqnarray*}
\int_{- \infty}^0 dx \int_{-\infty}^{\infty} dy ~
\big[ - (\partial_x \phi) (\partial_x \bar \phi) ~-~
(\partial_y \phi)(\partial_y \bar \phi)  && \\  ~+~
\coeff{i}{2}(\bar \lambda \gamma^\mu \partial_\mu \lambda -
(\partial_\mu \bar \lambda) \gamma^\mu  \lambda)) \big]
&&
\end{eqnarray*}
\begin{equation}
{} ~+~ \int_{-\infty}^{\infty}~  i \Big(b^\dagger {d \over dy} b
\Big) ~-~ \shalf  (\bar \lambda  \gamma^* \lambda)\big|_{x=0}~dy \ ,
\label{freeact}
\end{equation}
where $x = x^0$, $y = x^1$, $\bar \lambda$ is the hermitian
conjugate of $\lambda$, and
\begin{eqnarray*}
\gamma^0 = \left(\matrix{0 &i \cr-i &0 \cr} \right)  \qquad
\gamma^1 = \left(\matrix{0 &1 \cr 1 &0 \cr} \right) && \\
\qquad \gamma* = -i \gamma^0 \gamma^1 =
\left(\matrix{1 &0 \cr 0 &-1 \cr} \right) \ . && \\
\end{eqnarray*}
The boundary term in (\ref{freeact}) is motivated by the
boundary action for the Ising model (see, for example,
\cite{GZam}), and has the property that the variation of the
action cleanly enforces free or fixed boundary conditions on
the bulk fermion.

To the free action one adds bulk \LG superpotential terms:
\begin{eqnarray}
\int_{- \infty}^0 dx \int_{-\infty}^{\infty} dy~
\bigg[ \Big({\partial^2 W \over \partial \phi^2} \Big)~\lambda_1
\lambda_2   \qquad \qquad \ \qquad  && \nonumber  \\
{} ~-~ \Big({\partial^2 \overline W \over \partial\bar
\phi^2}  \Big)  ~\bar \lambda_1 \bar \lambda_2
{} ~-~\Big|{\partial W \over \partial \phi}  \Big|^2 \bigg] \ ,
&& \label{bulkpot}
\end{eqnarray}
for some scalar superpotential, $W$.

The bulk action is invariant, up to boundary terms,
under the following $N=2$ supersymmetry transformations:
\begin{eqnarray}
\delta \phi = -(\lambda_1 \bar \alpha_1 +
\lambda_2 \bar \alpha_2)\ , \quad \delta \bar \phi = (\bar
\lambda_1  \alpha_1 + \bar \lambda_2 \alpha_2)
\ ,  && \nonumber \\
 \delta \lambda_i = \epsilon^{ij} \Big[ -(\partial_x \phi
- i \partial_y \phi)~ \alpha_j + \Big({\partial
\overline W \over \partial\bar \phi} \Big)~ \bar \alpha_j
\Big] \ , && \nonumber  \\
\delta \bar \lambda_i = \epsilon^{ij} \Big[- (\partial_x
\bar \phi + i  \partial_y \bar \phi)~ \bar \alpha_j +
\Big({\partial W \over \partial \phi} \Big)~ \alpha_j \Big]
\ ,   &&
\label{susytrf}
\end{eqnarray}
where $\epsilon^{ij} = -\epsilon^{ij}$ and $\epsilon^{12} = +1$.

To have supersymmetry in the boundary theory, one must cancel, or
otherwise cause to vanish, all of the boundary terms in the
supersymmetry variation of (\ref{freeact}) and (\ref{bulkpot}).
One must also consider the ordinary variation of the action,
leading to the Euler-Lagrange  equations of motion, and
require that all the boundary terms generated in this variation
also vanish.
There are several ways to accomplish this, but if one wants
$N=2$ supersymmetry, and trivial boundary dynamics, then
there are two choices.  The first is to take
\begin{equation}
\alpha_1 = \alpha_2 = \alpha \ ; \qquad  \bar \alpha_1 =
\bar \alpha_2 = \bar \alpha \ ,
\label{cmplxsusy}
\end{equation}
along with Dirichlet boundary conditions:
\begin{equation}
\phi_{x=0} = \phi_0 \ ; \quad (\lambda_1 + \lambda_2)|_{x=0}
= 0 \ .
\label{dirbcs}
\end{equation}

The second choice results in a ``real'' form of
$N=2$ supersymmetry.  One must first add the
following to the boundary action:
\begin{equation}
\int_{-\infty}^{\infty}~(W(\phi) ~+~ \overline W
(\bar \phi))|_{x=0}  ~dy \ .
\label{bdryW}
\end{equation}
Then one takes:
\begin{eqnarray}
&\alpha_j ~=~ \bar \alpha_j \ ; \quad (\lambda_j + \bar \lambda_j)
|_{x=0} ~=~ 0 \ ;  \quad j = 1,2 \ ; & \nonumber \\
&(\phi - \bar \phi)|_{x=0} ~=~ 2i \phi_0 \ ; & \nonumber \\
& [\partial_x(\phi + \bar \phi)]|_{x=0} ~=~ \Big(
{\partial W \over  \partial \phi}  ~+~ {\partial \overline W \over
 \partial \bar \phi}\Big)\Big|_{x=0} \ . &
\end{eqnarray}

The foregoing identifications of the $\alpha_j$ and $\bar \alpha_j$
are simply a manifestation of the identifications given in
(\ref{Abcond}). The boundary term, (\ref{bdryW}), can be viewed as
the boundary part of the bulk topological charge. This term was
anticipated in section 3, and is needed to preserve supersymmetry
with the more general boundary conditions. However, this term is not
necessary for the Dirichlet conditions since such conditions are
sufficiently stringent to kill off the boundary terms arising from
the topological charge.

To obtain a theory with non-trivial boundary dynamics
(and with couplings of positive dimension), one can add the
boundary interaction:
\begin{eqnarray}
\shalf~\int_{-\infty}^\infty ~ \Big({\partial^2
V \over \partial \phi^2} \Big) ~ b^\dagger ~(\lambda_1 +
\lambda_2) ~-~ \qquad \qquad \quad  && \nonumber \\
\Big({\partial^2 \bar V \over \partial \bar
\phi^2} \Big)~b~(\bar \lambda_1 + \bar \lambda_2)  ~+~ \Big|
{\partial V \over \partial \phi} \Big|^2 ~dy\ , &&
\label{dbdryact}
\end{eqnarray}
where $V$ is some scalar potential for $\phi|_{x=0}$.

Consider the model consisting of (\ref{freeact}),
(\ref{bulkpot}) and (\ref{dbdryact}) (but {\it not} (\ref{bdryW})).
This model has $N=2$ supersymmetry provided one imposes
(\ref{cmplxsusy}) and
\begin{equation}
\Big( {\partial V \over  \partial \phi} \Big)
\Big|_{x=0} ~=~\mu~ W|_{x=0} \ .
\label{VandW}
\end{equation}
The boundary fermion fields transform according to:
\begin{eqnarray}
\delta b ~=~ 2 \bar \mu^{-1}\bar \alpha ~-~ \mu~
\alpha ~W|_{x=0}\ ; && \nonumber \\
\delta b^\dagger ~=~ 2 \mu^{-1}
\alpha ~-~ \bar \mu~\bar \alpha~\overline W|_{x=0}~  \ . &&
\label{deltab}
\end{eqnarray}
The parameter $\mu$ is an arbitrary complex number whose modulus
can be thought of as the ratio of the boundary and bulk mass scales.
The real and imaginary parts of $\mu$, along with the bulk mass
scale are precisely the three independent parameters identified
in the conformal perturbation theory of section 4.

It is interesting to observe that $N=2$ supersymmetry requires that
the boundary potential be fixed completely by the bulk potential, and
that if $W$ is a polynomial of degree $n$ then $V$ is a polynomial of
degree $n+1$. If one wants to have a theory in which the bulk and
boundary potentials are independent, then one can only have $N=1$
supersymmetry. Given the results of section 3, such independence is
also probably inconsistent with integrability. Having said this, one
should also note that more species of boundary fermions can be added,
with whatever potentials one desires, and one can still preserve
$N=2$ supersymmetry provided one of the boundary fermions has the
action described above. The choice of bosonic potentials for the
other fermion species is arbitrary, but only a small subset of these
models will be quantum integrable.

It is also important to observe that if $W$ is quasi-homogeneous:
that is, if $W(a^\omega \phi) = a W(\phi)$,
for some $\omega$ and any value of $a$, then the theory
has an $R$-symmetry.  Namely,
it is invariant under:
$\phi \to e^{i \omega \theta}\phi$,
$\bar \phi \to e^{-i \omega \theta}\bar \phi $,  $\lambda_j \to
e^{-{i \over 2} (1 - 2 \omega)  \theta}\lambda_j$, $\bar
\lambda_j \to e^{+{i \over2} (1 - 2 \omega) \theta}\bar \lambda_j$,
$(j = 1,2 )$, $\alpha \to e^{-{i \over 2} \theta}\alpha$,
$\bar \alpha \to e^{+{i \over2}\theta}\bar \alpha$,
$b \to e^{{i \over 2} \theta}b$,  $b^\dagger \to
e^{-{i \over 2} \theta}b^\dagger$, where $\theta$ is some parameter.
Such a symmetry is a critical element of establishing that
the quasi-homogeneous potential generates the superconformal
model.  In particular, this $R$-symmetry provides the $N=2$
superconformal $U(1)$ current.  This observation highlights
an, as yet, unresolved subtlety:  the form of the effective
action when the bulk is conformal but the boundary is massive.
If the bulk is conformal then $W$ is quasi-homogeneous, and
thus (\ref{VandW}) implies that the boundary potential must
be similarly scale invariant.  On the other hand, one knows that
the boundary  sine-Gordon model has independent bulk and boundary
masses, and has $W(\phi)= \phi^3 + \phi$.  Presumably the
resolution of this lies in taking the careful massless limit of
the bulk massive model.

\section{EXACT SCATTERING MATRICES}

The $N=2$ supersymmetric integrable models arising from perturbations
of the $N=2$ superconformal minimal series are fairly well
understood.  The bulk $S$-matrices are known, and some TBA
computations have been performed \cite{FMVW,PFKI,LNW,CGGS}.
In these models, the supersymmetry genertors are realized
as charges that commute with the $S$-matrix, and these
charges can be interpretted as a part of an affine
quantum group symmetry of the model.
It is thus natural to ask about the exact boundary reflection
matrices and their behaviour with respect to the supersymmetry.
Unfortunately, very few of the appropriate boundary reflection
matrices are known:  Of the $N=2$ supersymmetric models, only
the boundary sine-Gordon model has been studied in sufficient
detail, and so I will focus on this.  As will be seen, this
model is sufficient to highlight some interesting open problems.
A relatively detailed discussion may be found in \cite{NPW}, and
here I will merely summarize the ideas and conclusions.

The bulk $S$-matrix for sine-Gordon has been known for
some time \cite{ZandZ}, and as explained in \cite{ABL}, this
$S$-matrix commutes with the generators of
$U_q(\widehat {SU(2)})$, the affine quantum group based upon
$SU(2)$.  At the supersymmetry point, $\beta^2 = 16\pi/3$ one
has $q^2 = -1$, and the four quantum group generators
satisfy the fermionic anti-commutation relations of the
massive $N=2$ supersymmetry algebra.

The boundary is incorporated by finding an exact boundary reflection
matrix, which gives the amplitudes for solitons and anti-solitons to
reflect into one another. As described in
\cite{IVC,ESkly,MezNep,GZam,SPaulo}, such matrices
can be determined by boundary analogues of the Yang-Baxter equations,
crossing, unitarity and the bootstrap. For supersymmetry to persist
in the presence of the boundary, one seeks linear combinations of the
generators that commute appropriately with the reflection matrix. One
finds that for general boundary conditions (\ie general $M$, $m$, and
$\Phi_0$ in (\ref{bdrySG})), there is only {\it one} supersymmetry
remaining, and not two as one might expect from the analysis above.
On the other hand, two supersymmetries survive at two special
surfaces
in the parameter space: one corresponds to Dirichlet boundary
conditions while the other is like a free boundary
condition\footnote{This boundary condition has the property that
solitons always reflect as anti-solitons, and vice-versa. This
situation corresponds to free boundary conditions only when the bulk
is massless \cite{GZam,NPW}.}. At these two surfaces in parameter
space, the two $N=2$ superalgebras
differ in the same manner as the two superalgebras described in
section 5. The fact that one does not get $N=2$ supersymmetry, except
at special points, can be tracked directly to the problems of bulk
topological charges.

The resolution of the conflict between this result and the analysis
above is relatively straightforward.  One of the assumptions
in the derivation of the boundary reflection matrix in \cite{GZam}
is that the boundary has no structure, and in particular, can store
no charge.  As we saw, the construction of supersymmetric actions
explicitly required the introduction of boundary fermions and
non-trivial boundary dynamics.  Thus to get the correct boundary
reflection matrix one will have to allow for this possibility.

There are straightforward ways of decorating the boundary
reflection matrix of \cite{GZam}: One can formally glue a
particle to the boundary \cite{GZam,Ghosh}.
That is, one thinks of the new boundary as a combination
of the known boundary and a particle of formal rapidity, $\zeta$,
running parallel to the boundary.  The rapidity
of the boundary particle is formal since this particle is never
considered as hitting the boundary -- the particle is simply
``glued'' to the boundary forever.  The new boundary reflection
matrix, $\widetilde R$, has the schematic form:
\begin{equation}
\widetilde R (\theta) ~=~ S(\theta - \zeta) ~R (\theta) ~
S(\theta + \zeta) \ ,
\end{equation}
where $S$ is the bulk $S$-matrix and $R$ is the original reflection
matrix. The matrix, $\widetilde R$, satisfies boundary Yang-Baxter,
unitarity, crossing and bootstrap as a consequence of the fact that
$S$ and $R$ satisfy such conditions.

The new boundary reflection matrix inherits the structure
of the states of the boundary particle, and also has a free
parameter in the rapidity, $\zeta$.  If one chooses the parameters
in $R$ so as to preserve two supersymmetries, then the new
boundary reflection matrix will also preserve two supersymmetries
since the $S$-matrices basically commute with the
supercharges\footnote{There are some subtleties here concerning the
co-product, see \cite{NPW}.}.  Thus one can easily construct
$N=2$ supersymmetry preserving boundary reflection matrices,
and these matrices have an independent boundary scale parameter,
$\zeta$.

One can easily imagine gluing more and more particles to the
boundary, and getting ever more exotic boundary reflection
matrices.  In the field theory, this would correspond to
adding more and more boundary fermions, and in terms of
physics it would correspond to coupling the boson
to impurities of higher and higher spin.

Thus, there are natural conjectures for the boundary reflection
matrices of the $N=2$ supersymmetric boundary integrable models.
These conjectures still need to be tested via Bethe Ansatz since
there may be some other way of decorating the known boundary
reflection matrices, or some other more general solution to
the boundary Yang-Baxter, crossing, unitarity and bootstrap
conditions.

\section{FINAL COMMENTS}

Conformal perturbation theory, along with the results of
exact $S$-matrix theory provide a fairly compelling body of
evidence that there are $N=2$ supersymmetric, boundary
integrable models, and that these models should have
independent boundary and bulk mass scales.  There remain
some obvious issues to be investigated.  First, it would be
useful to know if the $N=2$ supersymmetry preserving, boundary
reflection matrices described here are the only such matrices
for the sine-Gordon model at the supersymmetry point.  In
addition, some TBA computations need to be done to check the
correspondence between the boundary actions and the
boundary reflection matrices.  For example, one could
compute the boundary entropy and check it against the
semi-classical arguments based upon the form of the boundary
potential, as in \cite{PFKondo}.

More generally, it would be valuable to have more examples of
$N=2$ supersymmetry preserving boundary reflection matrices.
The obvious models to investigate are those based upon the
Chebyshev perturbations of the higher members of the $N=2$
superconformal minimal series.

Finally, one has seen that the effective \LG field theory of
the bulk extends naturally to the boundary theory, and
already gives some semi-classical insight into the structure
of the quantum integrable model.  It remains to be seen how
much further this can be taken, and whether one can get as
much quantum exact information from the boundary \LG theory
as one gets in the bulk theory from knowledge of the bulk
\LG potential.

\newpage
\bigskip
\leftline{\bf Acknowledgements}
\medskip
I would like to thank P.~Dorey, J.~McCarthy, J.-B.~Zuber and
particularly P.~Fendley and H.~Saleur for valuable discussions on
this work. I am also grateful to the the University of Adelaide, the
Service de Physique Th\'eorique at Saclay and to the High Energy
Physics Laboratory of the University of Paris VI in Jussieu for
hospitality while parts of this work were done. I would also like
to thank the ICTP, and the organizers of this workshop for their
hospitality and the opportunity to present this paper.


\end{document}